\documentclass[aps,prc,preprint,doublespaced,superscriptaddress,nofootinbib,showpacs,showkeys]{revtex4}

\usepackage{graphicx}

\begin{document}


\title{Emulsion Cloud Chamber technique to measure the fragmentation of a high-energy carbon beam}        

\newcommand{\authorunina}
{Dipartimento di Fisica, Universit\`a di Napoli, via Cintia, 80126 Napoli Italy}

\newcommand{\authorinfnna}
{INFN Sezione di Napoli, via Cintia, 80126 Napoli Italy}

\newcommand{\authorjst}
{CREST, Japan Science and Technology Agency, Kawaguchi 332-0012, Japan}

\newcommand{\authorkek}
{High Energy Accelerator Research Organization (KEK), Tsukuba 305-0801, Japan}

\newcommand{\authornagoya}{Nagoya University, Nagoya 464-8602, Japan}

\newcommand{\authornaruto}{Naruto University of Education, Naruto 772-8502, Japan}

\newcommand{\authornirs}
{Research Center for Charged Particle Therapy, National Institute of Radiological Sciences, 4-9-1 Anagawa, Inage-ku, Chiba 263-8555, Japan}

\newcommand{\authornirso}
{Fundamental Technology Center, National Institute of Radiological Sciences, 4-9-1 Anagawa, Inage-ku, Chiba 263-8555, Japan}

\author{G. De Lellis\footnote{Corresponding author: giovanni.de.lellis@cern.ch}}
\affiliation{\authorunina}
\affiliation{\authorinfnna}

\author{S. Buontempo}
\affiliation{\authorinfnna}

\author{F. Di Capua}
\affiliation{\authorinfnna}

\author{A. Marotta}
\affiliation{\authorinfnna}

\author{P. Migliozzi}
\affiliation{\authorinfnna}

\author{Y. Petukhov\footnote{on leave of absence from Joint Institute for Nuclear Research, Dubna, Russia}}
\affiliation{\authorunina}

\author{C. Pistillo\footnote{now at University of Bern, CH-3012, Bern, Switzerland}}
\affiliation{\authorunina}

\author{A. Russo}
\affiliation{\authorunina}
\affiliation{\authorinfnna}

\author{L. Scotto Lavina}
\affiliation{\authorunina}
\affiliation{\authorinfnna}

\author{P. Strolin}
\affiliation{\authorunina}
\affiliation{\authorinfnna}

\author{V. Tioukov}
\affiliation{\authorinfnna}

\author{T.~Toshito}
\affiliation{\authorjst}
\affiliation{\authorkek}

\author{A. Ariga}
\affiliation{\authornagoya}

\author{N. Naganawa}
\affiliation{\authornagoya}

\author{Y. Furusawa}
\affiliation{\authornirs}

\author{N. Yasuda}
\affiliation{\authornirso}

\begin{abstract}
Beams of Carbon nuclei are used or planned to be used in various centers for cancer treatment around the world because of their therapeutic advantages over proton beams. The knowledge of the fragmentation of Carbon nuclei when they interact with the human body is important to evaluate the spatial profile of their energy deposition in the tissues, hence the damage to the tissues neighboring the tumor. In this respect, the identification of the fragmentation products is a key element. We present in this paper the charge measurement of about 3000 fragments produced by the interaction of $^{12}$C nuclei with an energy of 400 MeV/nucleon in a detector simulating the density of the human body. The nuclear emulsion technique is used, by means of the so-called Emulsion Cloud Chamber. In order to achieve the large dynamical range required for the charge measurement, the recently developed techniques of the emulsion controlled fading are used. The nuclear emulsions are inspected using fast automated microscopes recently developed. A charge assignment efficiency of more than 99\% is achieved. The separation of Hydrogen, Helium, Lithium, Berillium, Boron and Carbon can be achieved at two standard deviations or considerably more, according to the track length available for the measurement.
\end{abstract}



\maketitle

\section{Introduction}

Unlike the electromagnetic radiation conventionally used for cancer treatment, charged hadrons (charged particles experiencing nuclear interactions, e.g.~protons) deposit most of their energy in a restricted domain around the end of their ionization range. This leads to a high therapeutic effectiveness with minimal damage to neighboring tissues. This is the merit of the so-called hadron-therapy. The tissue thickness traversed before depositing their energy can be tuned by changing the energy of the nuclei, which is typically of a few hundreds MeV/nucleon. 

Among hadrons, the use of nuclei heavier than protons is expected to improve the therapeutic effectiveness. Light nuclei and in particular Carbon nuclei are now used or planned to be used for cancer therapy in a number of dedicated facilities around the world (see for instance Ref.~\cite{scholz} and references therein). 

The nuclear fragments generated in the interaction of the projectile nuclei inside the patient body go, however, beyond their ionization range thus producing some damage in the tissues downstream of the tumor. The study of the fragmentation of the projectile nuclei is therefore important to improve the precision achievable in hitting the tumor with minimal effects on the neighboring tissues. Moreover, the data obtained in fragmentation studies contribute to define the parameters of nuclear interaction models entering in computer programs which simulate the biological effects for the optimization of their effectiveness.

Studies of the nuclear fragmentation are currently carried out using a target which, from the point of view of nuclear interactions, has properties close to those of the human body. The nuclear fragments are conventionally observed by detectors external to the target. Their identification can be achieved by comparing the energy loss in a plastic scintillator and the total residual energy in a BGO scintillator~\cite{matsufuji,matsu1}. The total charge-changing and partial cross-section of $^{12}$C in carbon, paraffin and water was reported in Ref.~\cite{golov} by using etched track detectors (Cr-39). Measurements of $^{10}$B and heavier ions cross-section with several different targets were performed at GSI via the energy loss measurement in a large-area ionization chamber~\cite{schall}. 

The Emulsion Cloud Chamber (ECC) technique~\cite{ECC} consists of using a sequence of nuclear emulsion films interleaved with passive material and thus allows to integrate the target and the fragment detector in a very compact set-up. 
Nuclear emulsions allow the measurement of the fragments' emission angles event by event with granularity and space resolution at the micro-metric level.
The development of techniques of controlled fading of particle tracks~\cite{operaem} has opened the way to measurements of the specific ionization over a very broad dynamic range. 

The capability of identifying the fragment's electrical charge has been demonstrated by exposing a sequence of  emulsion films to beams of different nuclei~\cite{toshito}. In parallel to the work presented here, the fragmentation was studied with the ECC technique and very recently the total and partial charge-changing cross-sections of 200 - 400 MeV/nucleon carbon in water and polycarbonate were measured~\cite{toshi3}.    

This work is devoted to the detailed study of the ECC charge identification capabilities by using a Carbon beam and observing the nuclear fragmentation products. With respect to the work done in Refs.~\cite{toshi3,toshi2}, different procedures of controlled fading were adopted and different emulsion scanning and analysis techniques were used. In particular, we have adopted lower values for the fading temperature: decreasing the temperature has the drawback of making the charge identification more challenging and the merit of reducing the so-called fog (random background grains in the emulsion gelatin due to thermal excitation), thus improving the purity of the tracking. The procedure followed in this work has thus the merit of achieving a larger purity in the tracking and vertex reconstruction. Unlike the automated scanning system used in Refs.~\cite{toshi3,toshi2}, the one used in this work allows an efficient reconstruction of tracks with angles larger than 500 mrad, especially relevant for the identification of light fragments (Hydrogen and Helium).

We have constructed an ECC with polycarbonate as passive material, approaching the nuclear properties of the human body. The ECC was exposed to a beam of $^{12}$C nuclei with an energy of 400 MeV/nucleon. This energy is meant for deep seated tumors. The ionization of about 3000 nuclear fragments produced in their interactions in the ECC has been measured. In this paper we report on the fragment identification efficiency and on the purity obtained in this identification. The study of the fragmentation itself will be the subject of a forthcoming paper. We expect that the results achieved with this technique will be used as input by simulation programs and will be relevant for the optimization of the treatment planning, thus improving the effectiveness of the therapy.  

\section{Nuclear emulsions}

Nuclear emulsions~\cite{powell} provide tracking of charged particle trajectories with a very high spatial resolution in the three dimensions, with the capability of measuring their ionization.
The passage of ionizing particles produces a latent image which is turned into a sequence of silver $grains$ ($\sim$ 0.6 $\mu$m diameter) after a complex chemical process known as development. The grains
lie all along the trajectory of the particle which can thus be measured with
a sub-micrometer accuracy.

The use of nuclear emulsions as a tridimensional tracking device for particle physics was limited in the past by the lengthy scanning procedure, originally by visual inspection at the microscope. It has recently undergone a revival due to the impressive and rapid development of fast automated scanning systems~\cite{suts,ess,ess1}, with a scanning speed several orders of magnitudes higher than the first automated systems.  
A system devoted to very precise measurements was also 
developed~\cite{precision}: the trajectory of tracks is measured with an
accuracy of 0.06 $\mu$m in position and 0.4 mrad in angle, compared to about 2 mrad of systems oriented toward achieving the highest speed~\cite{suts,ess,ess1}. 

In nuclear emulsions, the trajectories of particles at the minimum of their ionizing power (MIPs) are observed
as thin tracks with a grain density of about 30 grains/100~$\mu$m. The
grain density in emulsion is almost proportional to the
energy loss and its mean rate is given by the Bethe-Bloch
equation~\cite{powell}. 

The proportionality of the grain density to the energy loss by ionization holds over a limited range, above which a saturation effect dominates. This effect 
prevents the charge measurement for high ionizing particles. Nevertheless, 
by keeping the emulsions for an appropriate time at a relatively high temperature (above 30$^\circ$ Celsius) and a high relative humidity (around 98\%), a fading\footnote{Progressive oxidation of the latent image centers, enhanced by high umidity and temperature conditions.} is induced which partially or totally erases the tracks of particles~\cite{operaem}. Thus, for instance, by a controlled fading films may be made unsensitive to MIPs and suited for highly ionising particles. This treatment is called refreshing. The combination of several
films having undergone different refreshing treatments after exposure allows to disentangle particles with different charge and thus largely different ionization, overcoming saturation effects. This is the basis of the method reported in Ref.~\cite{toshito}. 

We have used refreshing temperatures up to 38$^\circ$ Celsius. The choice of the adopted values for the refreshing parameters was obtained after different optimization trials and come from the request of keeping reasonably low the fog level (less than 8 grains/(10$\mu$m)$^3$) and achieving a good erasing rate. This optimization was performed by applying the different refreshing conditions to a few samples of films previously exposed to carbon ion beams.  The choice of lower values with respect to Ref.~\cite{toshi3} was motivated by the wish of keeping the fog values lower, thus improving the tracking purity.  

The emulsions used for this work belong to the same batch of films used in the OPERA experiment for the study of neutrino oscillations~\cite{opera}. The OPERA emulsion films~\cite{operaem} have been developed for large-scale, high precision experiments This development has allowed the first industrial production of nuclear emulsions (as for X-ray films), leading a considerably lower unit cost than previously possible though maintaining a  sensitivity similar to that of handmade films. The OPERA nuclear emulsions have been produced by the Fuji company\footnote{Fuji Film, Minamiashigara, 250-0193, Japan.}. They consist of 44~$\mu$m thick emulsion layers deposited on both sides of a 205~$\mu$m thick plastic support. The surface size of the emulsion films is 12.5x10.0 cm$^2$.  

\section{The Emulsion Cloud Chamber and the beam exposure}

We have built an ECC made of a sequence of nuclear emulsion films interleaved, as passive material, with 1 mm thick plates of polycarbonate (Lexan)
 in a multiple sandwich structure. The polycarbonate has a density of 1.15 g/$cm^3$ and an electron density of 3.6x10$^{23}$ cm$^{-3}$. For comparison, the electron density of water (the main constituent of the human body) is 3.3x10$^{23}$ cm$^{-3}$ while the polymethil methacrylate target used in Ref.~\cite{matsufuji,matsu1} has a density of 1.19 g/$cm^3$ and an electron density of 3.7x10$^{23}$ cm$^{-3}$. Thus the target has similar characteristics to water as far as nuclear interactions are concerned. 

\begin{figure}[htb]
\includegraphics[width=17pc,scale=0.4,angle=0]{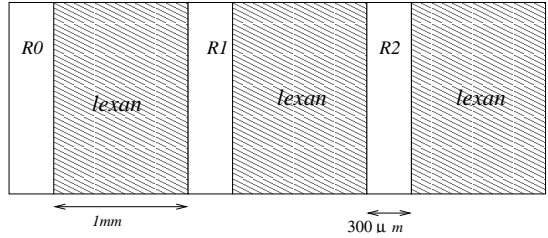}
\caption{Structure of one of the 73 cells of the Emulsion Cloud Chamber. Each cell is made of 3 different emulsion films interleaved by Lexan sheets.} \label{fig:exposure}
\end{figure}

The ECC consisted of 73 consecutive modules, each made of three emulsion films interleaved with polycarbonate plates. Its structure is shown in Fig.~\ref{fig:exposure}. It is significantly different from the one used in Ref.~\cite{toshi3} and the chamber length allows the study of the Bragg peak too. The 3 emulsion films, denoted as $R_0$, $R_1$ and $R_2$, were treated differently after the exposure and before their chemical development. $R_0$ was not refreshed and was developed soon after the exposure. $R_1$ and $R_2$ 
underwent a 3 day refreshing at 98\% relative humidity with 30$^{\circ}$C and 38$^{\circ}$C temperature,
respectively. The pile of emulsions and polycarbonate plates was vacuum packed. A light-tight aluminum tape was used to protect the pile from light.

\begin{figure}[htb]
\includegraphics[width=17pc,scale=0.4,angle=0]{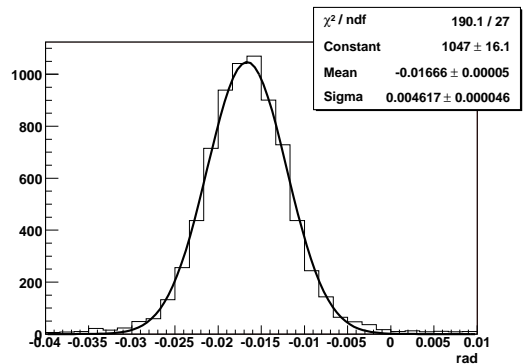}
\caption{Angular distribution of the incident carbon beam (rad).} \label{fig:beamangle}
\end{figure}

The ECC was exposed to a beam of $^1$$^2$C nuclei with an energy of 400 MeV/nucleon at the Heavy Ion Medical Accelerator (HIMAC) in Chiba (Japan). The beam flux was monitored by a scintillator counter. On the emulsion films, an integrated flux of about 10 $^{12}$C nuclei/mm$^2$ was obtained. This is a compromise between the need of a large statistics and the need of avoiding the overlapping between close interactions. 
The angle of the incident beam with respect to the emulsion films had a spread of a few mrad as shown in Fig.~\ref{fig:beamangle}. The ECC was placed in three position: with emulsion films perpendicular to the beam and inclined of $\pm$150 mrad. The tilted exposures were meant to improve the film to film alignment. 

The chemical development was carried out at the Nagoya University. After the development, the films were brought to Naples University where they were analyzed by fast automated microscopes operating at a speed of 20 cm$^2$/hour with tracking efficiency larger than 90$\%$ and high purity ($\sim 2$ fake tracks/$cm^2$)~\cite{ess,ess1}.    

\section{Analysis methods}
The automated microscopes have a focal depth of a few~$\mu$m. By varying the focal plane over 20 levels along the depth of the emulsion layers, they gather a series of tomographic images which are read by a CMOS camera. A track is seen as a sequence of grains at different depths, each grain consisting of a cluster of pixels. Apart from saturation effects, the grain density along the particle path is proportional to the specific ionization. Therefore as a variable sensitive to the specific ionization, hence to the particle charge, we take the sum of the pixels of all the grains belonging to the track normalized to a given track length in the emulsions. This sum is called track volume.

In the data acquisition, the automated microscope reconstructs the so-called micro-tracks, a micro-track being a sequences of aligned grains in a 44~$\mu$m emulsion layer. The alignment, within errors, of two micro-tracks in a film gives a so-called base-track. Base-tracks are characterized by a higher angular precision than micro-tracks, because of the lever arm given by the thickness of the plastic base and because the grains close to the base do not suffer the distortions resulting from the chemical development.  
\begin{figure}[htb]
\includegraphics[width=17pc,scale=0.4,angle=0]{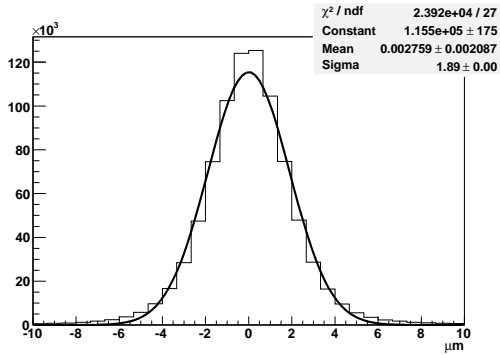}
\caption{Position resolution in the alignment of consecutive emulsion films ($\mu$m).} \label{fig:align}
\end{figure}
In the ECC, the emulsion films are piled up with a mechanical accuracy of a few hundred micrometers. By using straight penetrating beam tracks as references, in the offline analysis the films are aligned with an accuracy of a few micrometers~\cite{tracking} as shown in Fig.~\ref{fig:align}. After this fine alignment, base-tracks are associated to form tracks of particles.

In principle, each track is characterized by three track volume variables ($VR_0$, $VR_1$ and $VR_2$), one for each refreshing condition ($R_0$, $R_1$ and $R_2$) of the films traversed by the track. By averaging over the base-tracks in the emulsions having undergone the same refreshing, the statistical error on the track volume is reduced thus providing a better charge discrimination.

The $R_2$ refreshing procedure produces the complete erasing of all tracks of particle with charge equal to 1. Therefore, for proton identification only $VR_0$ and $VR_1$ are used. For Helium and heavier nuclei only $VR_1$ and $VR_2$ are effective, since $VR_0$ shows saturation. The charge separation is obtained by looking at correlations between appropriate pairs of track volume variables.

\section{Results}
All along the ECC, about 75\% of the incident carbon ions interact before showing the Bragg peak. The energy of the primary carbon ions used in this paper is 400 MeV/nucleon. Nevertheless, the longitudinal size of the analyzed volume is about 10 cm and therefore the carbon ions loose energy by ionization down to 200 MeV/nucleon. Thus the results shown in this paper concern all fragmentation products produced by the interactions of carbon ions in the energy range of 200$\div$400 MeV/nucleon.

\begin{figure}[htb]
\includegraphics[width=17pc,height=20pc,scale=0.8,angle=0]{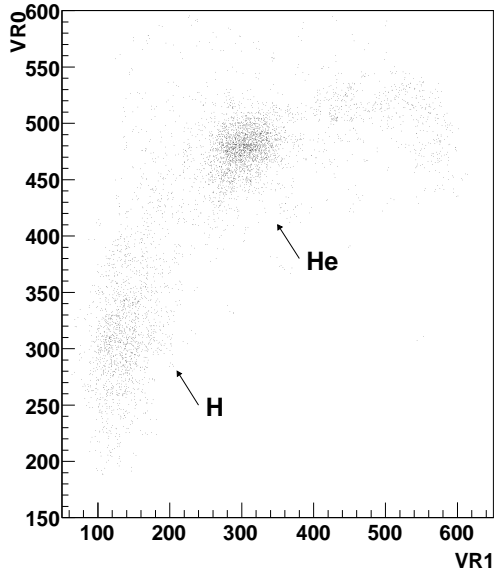}
\caption{Scatter plot of $VR_0$ versus $VR_1$, providing the separation of H from He and heavier nuclei.} \label{fig:R0R1}
\end{figure}

\begin{figure}[htb]
\includegraphics[width=17pc,height=20pc,scale=0.8,angle=0]{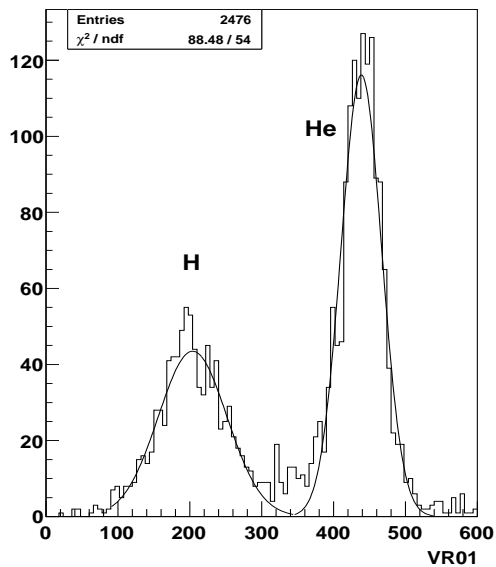}
\caption{The distribution of $VR_{01}$, providing the H and He separation. The tracks originate from the fragmentation of carbon nuclei all along the chamber. }
\label{fig:R0R1fit}
\end{figure}

\begin{figure}[htb]
\includegraphics[width=17pc,height=20pc,scale=0.8,angle=0]{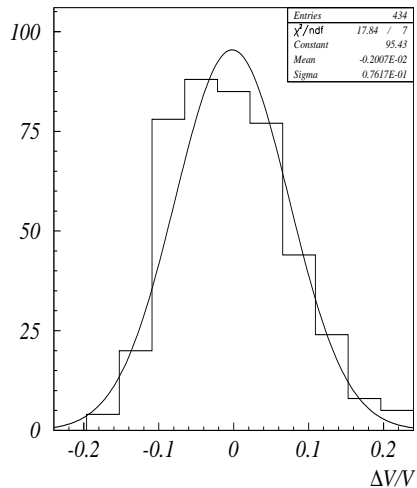}
\caption{The distribution of the relative difference in the ionization of the same particle as measured in two consecutive plates. }
\label{fig:ionization}
\end{figure}

Fig.~\ref{fig:R0R1} shows the scatter plot of $VR_0$ versus $VR_1$ for the fragmentation products. We see two distinct peaks corresponding to H and He. Heavier ions are not clearly identified. By projecting the scatter plot onto an axis passing through the centers of the two peaks, we obtain the distribution of the variable $VR_{01}$ shown in Fig.~\ref{fig:R0R1fit}. A good separation between H and He is visible. The events in-between the two Gaussian distributions are due to the wide momentum distribution of the H produced by the carbon nuclei interactions. This introduces a systematic uncertainty in the charge measurement at the level of a few percent: Fig.~\ref{fig:ionization} shows the relative difference of two ionization measurements performed in two consecutive plates. An uncertainty of about 7\% is measured which accounts also for the uncertainties in the development procedure. The use of the average value of the ionization improves the charge discrimination.  

\begin{figure}[htb]
\includegraphics[width=17pc,height=20pc,scale=0.8,angle=0]{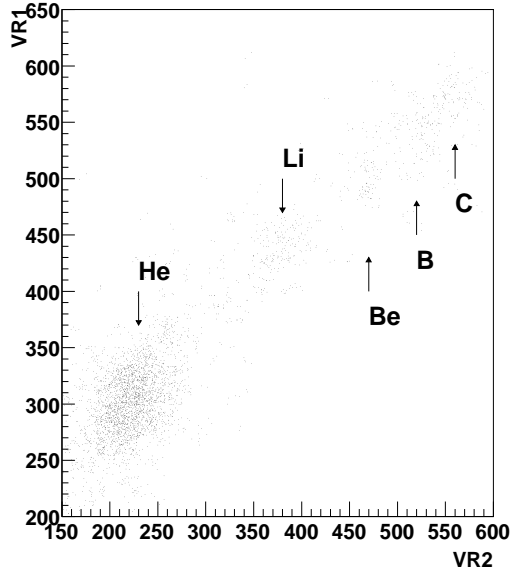}
\caption{Scatter plot of $VR_1$ versus $VR_2$. This provides the separation between He and heavier ions.} \label{fig:R1R2a}
\end{figure}

\begin{figure}[htb]
\includegraphics[width=17pc,height=20pc,scale=0.8,angle=0]{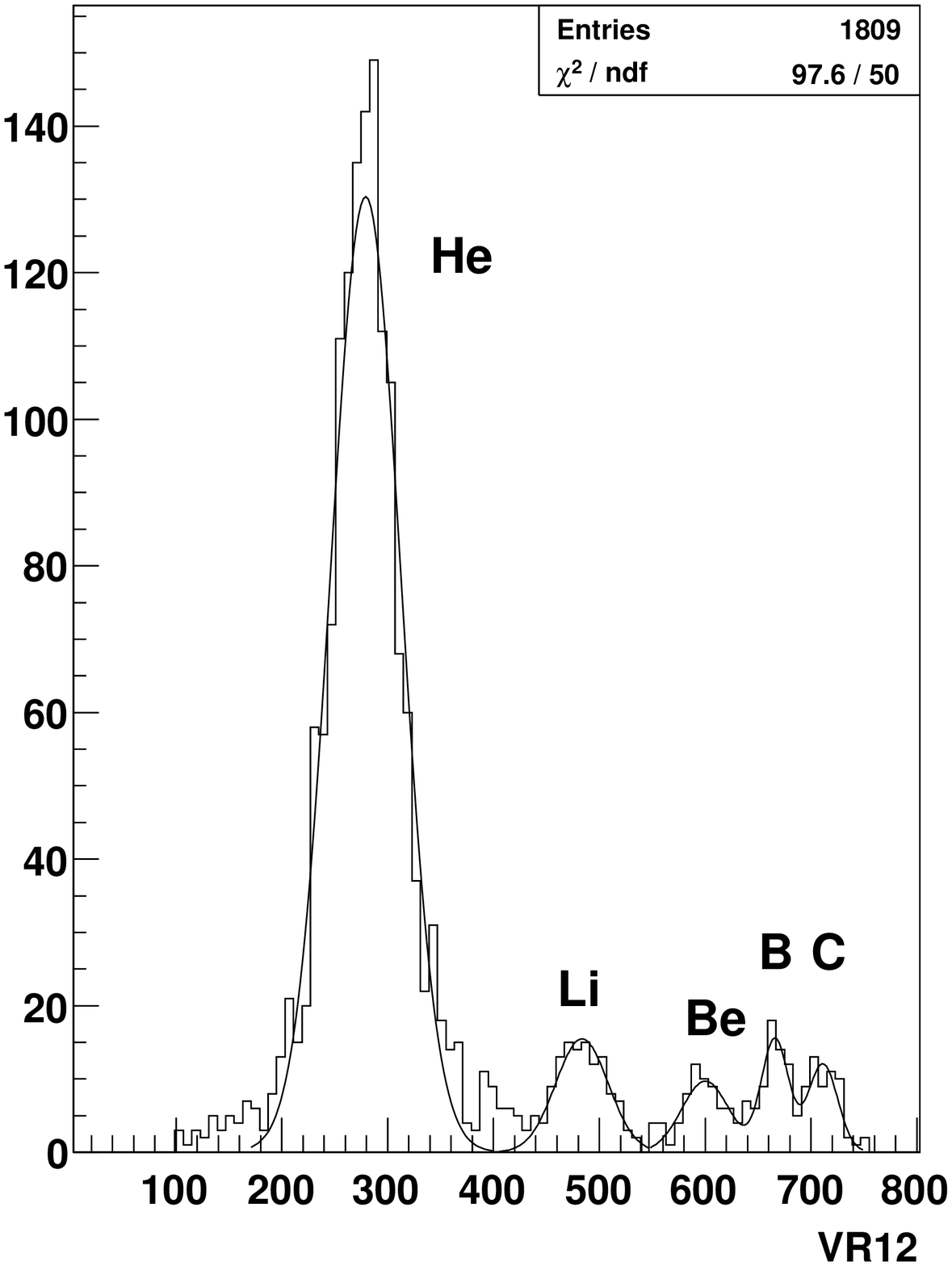}
\caption{The distribution of $VR_{12}$, showing the separation between He, Li, Be, B and C. The tracks originate from the fragmentation of carbon nuclei all along the chamber.}
\label{fig:R1R2_23fit}
\end{figure}

Fig.~\ref{fig:R1R2a} shows the scatter plot of $VR_1$ versus $VR_2$ for the fragmentation products. The separation of Helium, Lithium, Beryllium, Boron and Carbon becomes clear. The analysis reported here concerns the charge identification and therefore the different isotopes are not separated. By projecting the scatter plot onto an axis passing through the centers of the  peaks, we obtain the distribution of the variable $VR_{12}$ shown in Fig.~\ref{fig:R1R2_23fit}. Also in this case, the fraction of events in-between $He$ and $Li$ is due to the low momentum tail of the $He$ particles. The effect of the momentum spectrum becomes negligible for heavier nuclei since they tend to be produced with similar speed to the projectile.  

Table~\ref{table:sigma} gives the separation of pairs of nuclei in standard deviations, for a set of different values of the number of base-tracks. The standard deviation $\sigma$ for the separation of two nuclei is defined as $\sigma = \sqrt{\sigma^2_1+\sigma^2_2}$ where $\sigma_1$ and $\sigma_2$ are the standard deviations of the charge measurement for the two nuclei. 
In terms of total traversed ECC length, one base-track corresponds to 1.3 mm. One sees that to separate H-He, He-Li, Li-Be and Be-B at approximately 3 $\sigma$ one needs 3, 5, 12 and 30 base-tracks, respectively. With 30 base-tracks, Boron and Carbon have a 2.4 $\sigma$ separation. With sufficient track lengths, even better separations are achieved.

\begin{table}[htb]
\caption{The separation of pairs of nuclei in standard deviations, for a set of values of the number of base-tracks. Standard deviations are defined in the text. }
\begin{tabular}{|l|c|c|c|c|c|}
\hline
Base-tracks & 3 & 9 & 13 & 20 & 30 \\
\hline
H-He & 3.3 & 4.5 & 6.5 &  & \\
\hline
He-Li & 2.6 & 3.9 & 4.3 & 5.0 &  \\
\hline
Li-Be &  1.7 & 2.7 & 3.1 & 3.5 & 4.1 \\
\hline
Be-B &   &   & 2.0 & 2.5 & 2.8 \\
\hline
B-C &   &   &   & 1.9 & 2.4 \\ \hline
\end{tabular}
\label{table:sigma}
\end{table}

We have collected and analyzed about 3000 tracks. The charge assignment is done according to the ionization difference between the measured value and the average values of the different charge peaks: the closest value (in standard deviation units) is selected. The charge assignment failed only for 27 of them. The corresponding charge assignment efficiency is $(99.1\pm0.2)$\%. The inefficiency is connected to very short tracks. Indeed, the 27 tracks are tracks with only 2 base-tracks, in $R_0$ and $R_2$, thus preventing the estimate of both the variables $VR_{01}$ and $VR_{12}$.


\section{Conclusions}
We have exposed an Emulsion Cloud Chamber to a beam of $^{12}$C nuclei with an energy of 400 MeV/nucleon. The chamber was made of nuclear emulsion films and polycarbonate (Lexan) plates. We have analyzed about 3000 particle tracks produced by the Carbon fragmentation inside a given volume of the chamber. Although the incident carbon beam had a fixed energy, the analysis of interactions in the given volume has made the carbon energy ranging in the 200 to 400 MeV/nucleon energy range. By analyzing the grain density along the particle track by means of fast automated microscopes, we were able to assign the charge to these fragments with 99\% efficiency. The charge separation improves statistically with the track length. The separation of the Hydrogen, Helium, Lithium, Berillium, Boron and Carbon can be achieved at two standard deviations or considerably more, depending on the track length which is used for the measurement. 

\section*{Acknowledgments}
We thank Prof. M.~Durante for several suggestions and comments to the manuscript. We also thank Prof.~K.~Niwa and Prof.~G.~Gialanella for useful discussions and support. The Italian Ministry for University and Research (MUR) funded this work within the PRIN2004 projects. We acknowledge the Istituto Nazionale di Fisica Nucleare for the equipment used. In Japan, this work was performed as a Research Project with Heavy Ions at NIRS-HIMAC. We express our gratitude to the HIMAC laboratory and to its
staff for assistance.

\end{document}